\documentstyle[12pt]{article}
\begin{document}

\hfill CERN-TH/99-99

\medskip

\begin{center}
{\ {\Large {\bf THE INFLUENCE OF HIGH MULTIPLICITIES }}} \\
\bigskip
{\ {\Large {\bf AT RHIC 
{\footnote{ RHIC is an abbreviation of ``relativistic heavy-ion 
collisions''.}}
              ON THE GAMOV FACTOR }}}
 \\

\vskip 0.8cm

{\large {\sl D.V.~Anchishkin$^{\ a,\, b, \! \! }$
{\footnote{ E-mail: Dmitry.Anchishkin@cern.ch \ or/and \
                    Dmitry.Anchishkin@ap3.bitp.kiev.ua }},
W.A.~Zajc$^{\ c, \! \! }$
{\footnote{ E-mail: ZAJC@nevis1.nevis.columbia.edu }},
G.M.~Zinovjev$^{\ b, \! \! }$
{\footnote{ E-mail: GEZIN@ap3.bitp.kiev.ua }}
}}
\\
\vskip 0.80cm

$^{a}$CERN TH-Division, CH-1211 Geneva 23, Switzerland \\

\vskip 0.40cm

$^{b}$Bogolyubov Institute for Theoretical Physics, \\
 National Academy of Sciences of Ukraine \\
252143 Kiev-143, Ukraine\\

\vskip 0.40cm

$^{c}$Nevis Laboratories, Columbia University, \\
Irvington, NY 10533, USA\\

\vskip 0.40cm

\end{center}

\vskip 1.2cm

\begin{abstract}

The corrections for two-pion correlations due to electromagnetic
final-state interactions at high secondary multiplicities are investigated.
The analysis is performed by solving the Schr\"odinger equation
with a potential which is dictated by the multi-particle environment. 
Two different post-freeze-out scenarios are examined.
First, for a uniformly spread environment of secondary particles,
a screened Coulomb potential is exploited.
It is shown that the presence of a static and uniform post-freeze-out 
medium results in
a noticeable deviation from the standard Gamov factor.
However, after going to a more realistic model of an expanding pion 
system, this conclusion changes drastically.
We argue that the density of the secondary pions $n_{\pi}(t,R)$, where
$R$ is a distance from the fireball, 
is bounded from above by 
$n_{\pi}(t,R)\le {\rm const}/R^2$ for all times $t$.
Then, a two-particle scalar potential which is found as a solution of the 
Maxwell equation for non-uniform medium replaces the screened one. 
Even this upper limit does not result in an 
essential deviation from the Gamov correction.

\end{abstract}

\pagebreak

\section{Introduction}

Two-particle correlations provide information about the space-time
structure and dynamics of the emitting source \cite{boal}.
Considering the correlations that occur in relativistic
heavy-ion collisions one usually assumes that:
(i)  the particles are emitted independently (or the source is completely
     chaotic), and
(ii) finite multiplicity corrections can be neglected.
Then the correlations reflect a) the effects from symmetrization
(antisymmetrization) of the wave function and
b) the effects that are generated by the final-state interactions of the
detected particles between themselves and with the source.
At first sight one can regard the final-state interactions (FSI)
as a contamination of the `pure' particle correlations.
It should be pointed out, however, that the FSI depend on the structure of the
emitting source and thus provide information about the source dynamics as
well.
In fact, this was proved by intensive investigations during the last twenty years
of two-particle and source-particle FSI
\cite{boal,GKW} (for recent publications see, for example,
 on two-particle FSI, \cite{anch98,anch95} and references therein;
on source-particle FSI, \cite{barz96,barz97} and references therein).

Actually, the former approaches accounting for FSI deal with
the secondaries in empty post-freeze-out space.
Meanwhile, in recent SPS experiments, for instance ${\rm Pb+Pb}$ at
$160\ {\rm GeV}\times A$, some 800--900 secondary charged pions are created 
which form obviously a plasma-like post freeze-out medium.
Therefore, one might expect that the FSI of two separate 
pions, at this collision energy, and of course at the energies of the 
forthcoming RHIC and LHC colliders,
would be strongly influenced by the environment formed by other particles.
The goal of the present paper is to estimate how large are the
consequences on the Coulomb final-state interactions due to the presence of a 
large number of secondary charged particles.
To be as independent as
possible from source models, we chose the estimation of the Gamov
factor as a standard quantity, which serves as a measure of the Coulomb FSI.

The fundamental observable for intensity interferometry in hadron
physics is the relative momentum spectrum of identical particles.
For two like-sign pions, the modifications to the spectrum by the 
final-state Coulomb interaction result in a correction that has
typically been considered as tractable with high accuracy.
This assumption was based on the significantly different length
scales between strong  ($\propto 1/m_\pi $) and
Coulomb ($\propto 1/m_\pi \alpha$) interactions \cite{Sakharov,bf}
(here $m_\pi$ is the pion mass and $\alpha $ represents the fine structure
constant).
The correction may then be treated on the basis of the Schr\"{o}dinger
equation, resulting in the well-known
Gamov factor $G({\bf q})$
\begin{equation}
G(|{\bf q}|)=
\mid \psi_{\bf q} ({\bf r}=0)\mid ^2= \frac{2\pi \eta }{e^{2\pi \eta }-1}
\ ,
\label{eq10}
\end{equation}
where
$\psi_{\bf q} ({\bf r})$ is the two-particle wave function,
$\eta = \alpha m_{\pi }/|{\bf q}|$,
and ${\bf q}$ is the relative  momentum of the particles.

The nominal quantity expressing the correlation function in terms
of experimental distributions \cite{boal} is
\begin{equation}
C({\bf k}_1,{\bf k}_2)=
\frac{\displaystyle P_2\left({\bf k}_1, {\bf k}_2\right) }
{\displaystyle P_1\left({\bf k}_1\right) \,
P_1\left({\bf k}_2\right) }
\ ,
\label{eq1}
\end{equation}
\noindent where
$P_1\left({\bf k}\right) =E\, d^3N /d^3k$
and
$P_2\left({\bf k}_1, {\bf k}_2\right) = E_1 \, E_2 \ d^6N /(d^3k_1d^3k_2)$
are single- and two-particle cross-sections.
For point-like emitters it can be expressed 
(due to the factorization of the corresponding matrix element)
in terms of a product of the Gamov factor to the model
correlations \cite{GKW,anch98}:
\begin{equation}
C({\bf k}_1,{\bf k}_2)=G\left(|{\bf k}_1-{\bf k}_2|\right) \,
C_{\rm model}({\bf k}_1,{\bf k}_2)
\ .
\label{eq6.0}
\end{equation}
It should be mentioned that,
in the centre of mass of the pair the relative pion momentum
$|{\bf q}|=2|{\bf k}|$, with ${\bf k}={\bf k}_1=-{\bf k}_2$, coincides with invariant relative momentum
$q_{\rm inv}\equiv [(k_{1}+k_{2})^{2}-4m_{\pi }^{2}]^{1/2}$,
where $k_{1}$ and $k_2$ are the pion four-momenta in an arbitrary frame.

We consider corrections due to the Coulomb final-state 
interactions in a common scheme in which we do not take
into account the finite-size of the fireball (for a consistent treatment
of the source finite-size effects, see for example \cite{anch98}).
This means that we calculate
the correction factor in accordance with the formula
$G_{\rm corr}(|{\bf q}|)=\mid \psi _{\bf q}({\bf r}=0)\mid ^2$,
where now $\psi _{\bf q}(\bf r)$ is obtained from a numerical solution 
of the Schr\"odinger equation with the two-particle Coulomb potential 
{\it distorted by a multipion environment}.
Since, as shown in \cite{anch98}, the finite 
size of the emission source  softens the manifestation of the FSI, 
the `Gamov factor' tends to overestimate the FSI effects and therefore
it is the most sensitive quantity  
for deviations from the standard two-particle Coulomb interaction.

\section{Static multiparticle environment}

In the high multiplicity case,
when a post-freeze-out multipion environment cannot be neglected,
the relation between the two-particle
electromagnetic potential $\phi ({\bf r})$ and the local charge density is given by
\begin{equation}
 \nabla^2 \phi ({\bf r})=-4\pi e (n^{(+)}-n^{(-)})\ ,
\label{15}
\end{equation}
where $e=\sqrt{\alpha }$ is the elementary charge;
the density of charged pions $n^{(\pm )}$
is related  to  that of neutral pions $n^{(0)}$
via a Boltzmann factor:
\begin{equation}
n^{(\pm )}=n^{(0)}\, \exp{\left(\mp \frac{{\it e}\phi }{T_{\rm f}}\right)} \ ,
\label{16}
\end{equation}
The density $n^{(0)}$ of $\pi ^{0}$-mesons at
the freeze-out temperature $T_{\rm f}$
coincides with the equilibrium density of charged pions
in the absence of Coulomb interactions (we consider symmetrical nuclear 
matter).
In the limit
$ e\phi \ll T_{\rm f}$, Eq.(\ref{16}) can be rewritten as
\begin{equation}
n^{(\pm )}=n^{(0)}\, \left( 1 \mp \frac{{\it e}\phi }{T_{\rm f}}\right)\ 
\label{17}
\end{equation}
(this requires that the pions are not closer than $\sim10^{-2}$~fm to
one another at $T_{\rm f}\approx 200\ {\rm MeV}$), so that
\begin{equation}
\nabla^2 \phi ({\bf r})=\frac{4\pi  e^2}{T_{\rm f}} (    2n^{(0)}   )
\phi ({\bf r})\ .
\label{18}
\end{equation}
The solution of this equation is well known and given by a screened Coulomb potential
\begin{equation}
\phi _{\pi ^{\pm }}(r)=\pm {\it e}\frac{e^{-r/R_{\rm scr}}}{r}\ ,
\label{19}
\end{equation}
where
\begin{equation}
\frac{1}{R_{\rm scr}} =
\sqrt{\frac{8\pi }{3} \alpha } \cdot \sqrt{\frac{n_{\pi }}{T_{\rm f}} }
\label{20}
\end{equation}
with the total pion density $n_{\pi }=3\, n^{(0)}$.
Thus, for like-sign pions the potential energy reads
\begin{equation}
U_{\pi \pi }(r)=\frac{\alpha }{r}
\exp{\left( -\frac{r}{R_{\rm scr}} \right) } \ .
\label{20a}
\end{equation}

To evaluate the correction factor we use the
screened Coulomb potential (\ref{20a}) to solve  the
Schr\"{o}dinger equation numerically,
for the two choises:
1) $n_{\pi}=0.25\, {\rm fm}^{-3}$
and 2) $n_{\pi}=0.03\, {\rm fm}^{-3}$. 
(From now on we shall quote these two cases as ``{\it LHC}'' and 
``{\it SPS}'' freeze-out conditions, respectively.)
Taking $T_{\rm f}=190\ {\rm MeV}$, for example, we obtain from
Eq.~(\ref{20}) screening radii 
~$R_{\rm scr} \approx 7.9$ fm ({\it LHC}),
~$R_{\rm scr} \approx 22.4$ fm ({\it SPS}), respectively. 
For completeness of illustration we also consider the intermediate case
$R_{\rm scr}\approx 19.3\ {\rm fm}$.

The results of these calculations are plotted in Fig.~1
together with the standard Gamov factor.
We see that a substantial correction to the standard Gamov factor
would be required even for the existing experimental data if one
adopts the idealized picture of a uniform post-freeze-out 
density of the environment.

It is interesting to point out that
the correction factor $G_{\rm cor}(q)$, which was evaluated
for the same screened potential using the
quasi-classical approximation \cite{anch96}, does not make a big difference
with the correction factor in Fig.~1.

Actually, what we really learned from the consideration of this idealized
scenario 
is that a cancellation of the Coulomb potential tail (by screening)
results in an increase of the correction factor in the region of
small relative momentum ($q \le 50~\rm{MeV}$), as can be seen from Fig.~1.
Thus the long-distance behaviour of the potential is responsible for the
dramatic deviation of the correction factor from the standard Gamov factor
at small relative momenta.
On the other hand, the tail of the two-particle potential
gives the main contribution to interactions when the particles are 
at large distances from one another,
where in turn the density of secondary particles is small in the realistic 
picture.
To keep this qualitative speculation as a thread we next turn to a more
realistic calculation, explicitly incorporating expansion.

\section{Expansion scenario}

In the previous scenario the whole position space was filled by particles
with a constant density; this is certainly an unrealistic approximation
(idealization).
In order to take into account post-freeze-out expansion of the pion system,
we parametrise the pion density as
\begin{equation}
n(R)=n_{\rm f}\frac{R_{\rm f}^2}{R^2}\ ,
\label{24}
\end{equation}
where $n_{\rm f}$ is the freeze-out pion density 
and $R_{\rm f}$ 
is the freeze-out radius.
Indeed, the spatial volume of the expanding pion system in the solid angle
$\Omega $ increases as $\Delta V=\Omega \cdot R^2 \cdot \Delta R$,
where  $R$ is the distance from the centre of the fireball and $\Delta R$
is the thickness of the layer, which we keep constant.
Then, if the number of particles $\Delta N$ in this volume is constant,
the density reads:
$n(R)=\Delta N/(\Omega \cdot R^2 \cdot \Delta R)={\rm const}/R^2$.
Thus, the model (\ref{24}) implies that all particles have the same modulus
of radial velocity.
As we shall see further, such a model still means an overestimation of 
the particle density.

To support our assumption (\ref{24})
let us consider a classical pion phase-space distribution.
After freeze-out it satisfies the collisionless Boltzmann kinetic equation
\begin{equation}
\frac{\partial f(x,p)}{\partial x_0} + {\bf v}\cdot \nabla f(x,p) =0 \ ,
\label{1}
\end{equation}
where ${\bf v}={\bf p}/p_0$ is the velocity of the particle
and $p_0=\omega ({\bf p}) \equiv \sqrt{m^2_{\pi }+{\bf p}^2} $.
We look for the expansion solution of this equation, which can be fixed by
asymptotic condition, for instance
$ \lim_{t \to \infty} f(t, {\bf x}=0;{\bf p})=0$.
A solution of this type can be written in the form
\begin{equation}
f(t,{\bf R},{\bf p}) = f_{0}({\bf R}-{\bf v}t,{\bf p})\ ,
\label{104}
\end{equation}
where
$f_0({\bf R},{\bf p})$ is the initial distribution ($t=0$).
For the sake of simplicity, we take an isotropic initial distribution 
in position and momentum spaces:
just before freeze-out the
particles were distributed (a) in accordance with
Boltzmann's law in momentum space, and (b) in accordance with a Gaussian
distribution in position space.
The classical kinetic equation
is quite sufficient to describe the {\it collective} (!)
behaviour of the pion system after freeze-out.
Hence, we assume that the system, at time $t=0$, occupies the phase 
space according to the distribution function
\begin{equation}
f_0({\bf R},{\bf p}) = n_0({\bf R}) g_0({\bf p}) \ ,
\label{101}
\end{equation}
where
\begin{equation}
n_0({\bf R}) =
\frac{N_{\pi }}{\left( 2\pi R_{\rm f}^2   \right) ^{3/2}}
\exp{\left( -\frac{R^2}{2R_{\rm f}^2} \right)} 
\label{102}
\end{equation}
with
$\int d^3 R\, n_0 ({\bf R}) = N_{\pi }$,
and
\begin{equation}
g_0 ({\bf p}) =
\frac{\displaystyle 2\pi ^2}
{\displaystyle m^2 _{\pi } T_{\rm f} K_2\left( \frac{m}{T_{\rm f}} \right)}
\exp{\left( -\frac{\sqrt{m^2_{\pi }+{\bf p}^2}}{T_{\rm f}} \right)} 
\label{103}
\end{equation}
with
$\int [d^3 p/(2\pi )^3]\, g_0({\bf p}) = 1$.
Here $N_{\pi }$ is the total number of pions,
$T_{\rm f}$ and $R_{\rm f}$ are the temperature and the mean radius of the system
at time $t=0$ (freeze-out), respectively, $m_{\pi }$ is the pion
mass and $K_2$ is a Bessel function of imaginary argument.

The spatial distribution of the particles at time $t$ is determined by
integrating the distribution function (\ref{104}) over the momentum variable
\begin{equation}
n(t,{\bf R}) = \int \frac{d^3 p}{(2\pi )^3 } \,
n_0 \! \left({\bf R}- 
\frac{\small {\bf p}}{\small \omega ({\bf p})} t \right) g_0 ({\bf p}) 
\ .
\label{105}
\end{equation}
Because of the  spherical symmetry, it is reasonable
to look at the radial density of pions
\begin{equation}
n_{\rm sph} (t,R) \equiv 4\pi R^2 n(t,R)
\, .
\label{106}
\end{equation}
This quantity may be treated as the number of pions in the shell with
unit thickness at time $t$ and at a distance $R=|{\bf R}|$
from the fireball centre.
Hence, $n_{\rm sph} (t,R)$ is a one-dimensional spatial distribution 
function and,
evidently, the area under this curve at any time is equal to
the particle number $N_{\pi }$, because of its normalization
$\int _0^{\infty }dR n_{\rm sph} (t,R) =N_{\pi }$.
We evaluate this function in accordance with  Eq.~(\ref{105}) 
for {\it SPS} freeze-out conditions:
$n_{\rm f }=0.03\, {\rm fm}^{-1/3}$,
$T_{\rm f} = 190~{\rm MeV}$ and $R_{\rm f}=7.1~{\rm fm}$.
The results of this calculation at different times $t$ are given in Fig.~2.
It shows that the spherical distribution is always almost Gaussian-like 
and the velocity of the distribution maximum is very close to the 
velocity of light.
The horizontal line (see Fig.~2) denotes a constant
spherical density $4\pi R^2 n(R)={\rm const}$.
This line is nothing more than the 3-dimensional spatial density 
$n(R)={\rm const'}/R^2$
(for the {\it SPS} freeze-out conditions ${\rm const}' \approx 85/4\pi $).
It follows that, at any time $t$,  
for the post-freeze-out particle density $n(t,R)$ 
the  upper limit is valid
\begin{equation}
n(t,R) \le
n_{\rm f} \frac{R_{\rm f}^2}{R^2} 
\ ,
\label{106a}
\end{equation}
where $n_{\rm f}={\rm max}\{ n(t=0,R)\} $ 
and the equality is reached for an expanding system where the
radial particle velocities are equal. 
Because of momentum dispersion in the post-freeze-out pion system,
the pion density
(\ref{24}) is an {\it essential overestimation} of the real density, which
is formed by the comoving multipion environment.
Hence, adopting the stationary density dependence (\ref{24}) we 
overestimate the 
influence of the environment, and  consequently overestimate a
distortion of the Coulomb FSI in the expansion scenario.
On the other hand, this approach provides us with a possibility to 
consider a stationary post-freeze-out environment even in the frame of 
the non-stationary expansion scenario (see Fig.~2).

Adopting the parametrization (\ref{24}), 
we get Eq.~(\ref{18}), where the pion density $n_{\pi }$ now depends on $R$:
\begin{equation}
 \nabla ^2  \phi ({\bf r})=
\frac{8\pi \alpha }{3\, T_{\rm f}} n(R)\, \phi ({\bf r})\ ,
\label{51}
\end{equation}
where we put $n^{(0)} \approx n_{\pi }/3$ as before
($n_{\rm f}=n_\pi$).
Inserting the pair centre of mass distance $R$ together 
with the distance $r$ between two detected particles, we have, in the classical
approach: 
\begin{equation}
R \approx R_{\rm f} + v_{\rm cm}\cdot t \ ,
\hspace{1.5cm} r \approx v_{\rm rel}\cdot t \ ,
\label{26}
\end{equation}
where $v_{\rm cm}$ is the velocity of the two-particle centre of mass
in the fireball rest frame and $v_{\rm rel}$ is the relative velocity of the
particles ($v_{\rm rel}=q/m$, $t=0$ is fixed on the freeze-out hyper-surface).
Eliminating the time from the approximate equalities (\ref{26}) one has
\begin{equation}
R=R_{\rm f} + \frac{v_{\rm cm}}{v_{\rm rel}} \cdot r\ ,
\label{27}
\end{equation}
which means that we parametrize the time evolution by the mean distance $r$ 
between two detected particles.

We can thus rewrite the dependence of the pion density on $r$ as
\begin{equation}
n\big( R(r) \big) =\left(\frac{v_{\rm rel}}{v_{\rm cm}} \right)^2
\frac{n_{\rm f }R_{\rm f}^{2}}{(r+\overline{r} )^2}
\ ,
\label{27a}
\end{equation}
where we define the dynamical freeze-out radius
\begin{equation}
\overline{r}\equiv R_{\rm f} \frac{v_{\rm rel}}{v_{\rm cm}}  \ .
\label{27b}
\end{equation}

It is time to recapitulate what has been done. First, we eliminate the
time dependence of the particle density, bounding it from above
by a stationary non-uniform spatial distribution,
in accordance with inequality (\ref{106a}). Then, we connect the mean 
distance $R$ of the
pair c.m.s. from the fireball and the mean distance $r$ between  
pions, thus eliminating again the time dependence. 
This means that all time and $R$ dependences are now parametrized
by the variable $r$. If the relative velocity of the 
separate  pions is small, then the partial time derivative 
$\frac{\partial}{\partial t} =v_{\rm rel}\, \frac{\partial}{\partial r} $
can be neglected. Hence, the problem is approximately reduced  to a 
stationary one. Solving Eq.~(\ref{51}) with particle density from (\ref{27a}), 
one obtains the two-particle potential 
energy $U(r)=e\, \phi (r)$, which will then be exploited in the stationary
Schr\"odinger equation to find a wave function.
This will be the further strategy of our estimations.  
So, the problem of a time-dependent (expansion scenario) is reduced 
to a stationary one by the price of somewhat overestimating the 
Coulomb corrections.

Equation (\ref{51}) may be rewritten as
\begin{equation}
\nabla^2  \phi ({\bf r}) =
\frac{c^2(q)}{(r+\overline{r} )^2 } \, \phi ({\bf r})\ ,
\label{52}
\end{equation}
where we note explicitly that when the particles are separated by
large distances $r$ the density of the multiparticle environment
goes down in accordance with (\ref{27a}).
The quantity $c^{2}(q)$ is defined in the following way:
\begin{equation}
c^2 (q) = \frac{8\pi \alpha }{3} \frac{R_{\rm f}^{2}n_{\rm f }}{T_{\rm f}}
\frac{v_{\rm rel}^2}{v_{\rm cm}^2}\ .
\label{53}
\end{equation}
Fixing the screening radius at freeze-out 
\begin{equation}
R_{\rm scr}^f=
\sqrt{\frac{3T_{\rm f}}{8\pi \alpha n_{\rm f }} } \ ,
\label{53a}
\end{equation}
we have
\begin{equation}
c(q) =  \frac{R_{\rm f}}{R_{\rm scr}^f} \frac{v_{\rm rel}(q)}{v_{\rm cm}}
\ .
\label{53b}
\end{equation}
In spherical coordinates, Eq. (\ref{52}) can be rewritten as
\begin{equation}
 \frac{d^2\phi (r)}{dr^2} + \frac{2}{r} \frac{d\phi (r)}{dr} -
\frac{c^2(q)}{(r+\overline{r} )^2 } \, \phi (r)\, =\, 0
\ .
\label{54}
\end{equation}
This equation may be solved by
\begin{equation}
\phi (r)= \frac{e}{r} \left( \frac{\overline{r}}{r+\overline{r}} \right) ^b
\ ,
\label{55}
\end{equation}
where $\phi (r)$ satisfies the boundary condition:
$\phi (r) \to e/r$ when $r \to 0$.
The exponent $b$ is a solution of the quadratic
equation resulting from the substitution of the ansatz (\ref{55}) into
Eq.~(\ref{54}) and takes the form (assuming proper asymptotic behaviour of
the potential $\phi $):
\begin{equation}
b(q)=-\frac{1}{2} +\frac{1}{2}\sqrt{1+4c^{2}(q)}
\, .
\label{56a}
\end{equation}
One has $b \rightarrow  0$ when $n_{\rm f}  \rightarrow  0$ and hence the
potential $\phi (r)$ transforms into the Coulomb potential in this case.
It is shown in Figs.~3a and 3b for {\it LHC} and {\it SPS} freeze-out 
conditions, respectively.
In the interval of interest, $b$ increases with increasing relative 
velocity, and so do deviations from a pure Coulomb field.
This intriguing
behaviour is directly related to the ``Hubble-like'' expansion implied by
Eq.~(\ref{27}): it becomes clear if we remember that the
pion density decreases (hence $R_{\rm scr} \rightarrow \infty $)  with
increasing distance $R$.
The expansion thus results in modifications to the Coulomb
potential that are of power-law, not exponential, form, in contrast to
the static result given by Eq.~(\ref{19}).

The connection of the potential (\ref{55}) and screened one can be found 
in the following way.
One may treat the corrected potential obtained in Eq.~(\ref{55}) as
an effective charge distribution
\begin{equation}
e_{\rm eff}=e\left(\frac{\overline{r} }{r+\overline{r} } \right)^b
\ ,
\label{58}
\end{equation}
which we are going to average.
Equation (\ref{52}) can be rewritten as
$ (\nabla^2 - \kappa ^{2})\phi ({\bf r})=0$,
with
$\kappa \equiv c(q)/(r+\overline{r} ) $.
This is equivalent to an $r$-dependent screening radius, i.e.
\begin{equation}
R_{\rm scr}(r)=\frac{r+\overline{r} }{c(q)} \ .
\label{25}
\end{equation}
As shown in Figs.~3, the deviation of the potential (\ref{55})
from the pure Coulomb form in the region of small relative pion momentum
$q\le 30\ {\rm MeV}$, is small ($b \to 0$). 
Since, $\kappa \to 0$ when $v_{\rm rel} \ll 1$
(see Eq.~(\ref{53b})),  
we first ignore the $r$ dependence of $\kappa $
to obtain the solution for the electromagnetic $\pi \pi $ potential in
the form of Eq.~(\ref{19}), then substitute the $r$-dependent $\kappa $
into Eq.~(\ref{19}), to find that
\begin{equation} U_{\pi \pi } =
\frac{\alpha e^{-c(q)r/(r+\overline{r} )}}{r}
\label{59}
\end{equation}
no longer decreases exponentially with distance when $r$ is large enough 
$r\gg \overline{r} $.
Instead, the numerator on the r.h.s. of Eq.~(\ref{59}) represents the averaged
charge distribution (\ref{58}) squared
(we are now considering the potential energy $U_{\pi\pi} $ rather than the
electric potential $\phi$, hence the extra factor of charge leading
the $\alpha$).

It is clear from Eq.~(\ref{53b}) that if
$v_{\rm rel}/v_{\rm cm} \ll 1$ ($R_{\rm f}$ and $R_{\rm scr}^{f}$
are of the same order for high multiplicities) the renormalized
constant $\alpha _{\rm eff}=\alpha \exp{[-c(q)]} $ is close to the bare value of
$\alpha $. Moreover, the same qualitative result comes from
the $r$-behaviour of the screening radius (\ref{25}) when it approaches the
asymptotic value $R_{\rm scr}=\infty $ (Coulomb law) with increasing $r$.
The quantity $c(q)$ increases with relative pion momentum, leading to larger
deviations from the Coulomb potential and agreement with the features of
the potential (\ref{55}) (see discussion after Eq.~(\ref{56a})).

\subsection{Evaluations}

The numerical evaluations of the correction factor (Gamov factor) should 
be provided by the solution of the Schr\"{o}dinger equation 
with the potential
\begin{equation}
U_{\rm expan}(r) = \frac{\alpha }{r}
 \left( \frac{\overline{r} }{r+\overline{r} } \right) ^b
\ ,
\label{60}
\end{equation}
where the exponent $b$ is momentum-dependent, in accordance with (\ref{56a}).
Then, one can construct the correction factor
$G_{\rm cor}(|{\bf q}|)=\mid \psi_{\bf q}({\bf r}=0)\mid ^2$.
To compare it with the standard Gamov factor $G_0$ we calculate
the ratio $G_0/G_{\rm cor}$ for the LHC freeze-out conditions.
The results are depicted in Fig.~4 as dotted curves
for three different values of the pair mean momentum $P_{\rm cm}$.

 We now correct the two-particle potential (\ref{60})
for small distances $r\leq a\equiv n_{\rm f}^{-1/3}$,
where $n_{\rm f}$ is the freeze-out pion density,
i.e. in the region where the distance $r$ between two pions is smaller
than the mean distance between the particles in the gas.
Indeed, there is no `screening' effect for this distance, and hence the potential
should be the Coulomb one.
There is no such a problem for the statically screened potential
$U(r)=\alpha \exp{(-r/R_{\rm scr})} /r$,
because the screening radius obeys the condition $a\ll R_{\rm scr}$
(by definition in the sphere
$4\pi R_{\rm scr}^3/3$ the number of particles is much larger than 1);
for the region $r\leq a\ll R_{\rm scr}$, the potential $U(r)$ automatically
transforms into the Coulomb one, $U(r)=\alpha /r$.
In the expansion scenario we have in addition to
$R_{\rm scr}$ another scale parameter,
$\overline{r} =R_{\rm f}\left(v_{\rm rel}/v_{\rm cm}\right)$,
where $R_{\rm f}$ is the freeze-out size of the system
($R_{\rm f}=7-10~{\rm fm}$ for {\it SPS} and {\it LHC} freeze-out 
conditions).
By construction the potential $U_{\rm expan}(r)$  approaches
the Coulomb one if (i) $r \ll \overline{r} $ (see (\ref{60})).
On the other hand, the scale parameter $\overline{r} $
may be smaller than the mean distance between particles in the gas
(ii) $\overline{r} < a$. 
In Fig.~5 we depict $\overline{r}$ for different mean momenta of the pair:
$P_{\rm cm}=50,~150,~450~\rm{MeV/c}$  and for the mean distance between 
particles at freeze-out $ a= n_{\rm f}^{-1/3}$ (horizontal lines).
For instance, as is seen in Fig.~5, for $P_{\rm cm}=450~\rm{MeV/c}$
and relative momentum $q<60~\rm{MeV/c}$, we have $\overline{r}<a$ 
for {\it SPS} freeze-out conditions.
Hence, combining (i) and (ii) we get that the asymptotic regime, i.e. the 
Coulomb potential, can be achieved only for $r \ll  a$.
But we know that for separation distances which obey $r < a$, all distortions 
of the Coulomb potential vanish.
Thus, the behaviour of the potential (\ref{60}) must be corrected, since it
should be a Coulomb one at distances that are smaller than
the mean distance between particles.
To improve the behaviour of the potential for distances smaller than $a$,
we use a smooth `potential switcher'
\begin{equation}
s(x)= \frac{1}{2} - \frac{1}{2} {\rm tanh}\, [2(x-2)] \ ,
\label{21c}
\end{equation}
which is depicted in Fig.~6.
Then, incorporating the
correct behaviour of the potential at small distances, we finally obtain
the potential energy in the following form
\begin{equation}
U_{\pi \pi }(r) = s\left(\frac{r}{a} \right) \, \frac{\alpha }{r} +
\left[ 1-s\left( \frac{r}{a} \right) \right] \,
U_{\rm expan}(r)
\ .
\label{59b}
\end{equation}
Certainly, with increasing time the mean distance between particles
increases, and the Coulomb potential is thus switched on at
distances larger than the freeze-out particle mean distance 
$n_{\rm f}^{-1/3}$.
But we restrict our calculation to the freeze-out mean distance. 
It means that in the competition between the
two potentials $U_{\rm Coulomb}$ and $U_{\rm expan}$ we overestimate the 
contribution
of the distorted potential $U_{\rm expan}$ and hence we overestimate the
influence of the multiparticle environment.

The results of the calculation with potential
(\ref{59b}) are shown for the {\it LHC} freeze-out conditions
as solid curves in Fig.~4.
The same ratio $G_0/G_{\rm cor}$ for the {\it SPS} freeze-out conditions 
is given in
Fig.~7. The correction factor we obtained reveals only small deviations 
from the standard Gamov factor. 
This result can be explained by a fast decrease of 
the density of secondary particles with increasing distance of the pair 
from the fireball, which in turn results in a very small distortion of the 
two-particle Coulomb potential.

It is instructive to point out that evaluations that were made 
by the authors in \cite{anch96} on the 
basis of a quasi-classical approximation are in good qualitative agreement 
with the present results.

\section{Summary and conclusions}

In our first approach (static scenario) it was assumed that the whole 
position space is filled by secondary pions with  constant density.
This uniform environment of secondary pions results in
a screened two-pion Coulomb potential.
We showed that for future LHC and RHIC experiments the
screening radius of the Coulomb interaction at the freeze-out density
is of a size comparable with the source, and therefore the
factorization of Eq.~(\ref{eq6.0}) \cite{GKW} is no longer valid. 
Moreover, solutions of the Schr\"odinger equation produce a 
correction factor $G_{\rm cor}=|\psi_{\bf q}({\bf r}=0)|^2$ which  
noticeably deviates from the standard Gamov factor (see Fig.~1).
However, as we showed further, this model is a quite unrealistic 
approximation
and is not relevant to the real picture of an expanding pion system
after freeze-out.  

The conclusions reached with the first model change drastically 
after passaging to a more realistic model of an expanding pion 
system.
In the second model, we first reduce the time evolution of 
the multipion post-freeze-out environment to a stationary one,
parametrizing the density of the secondary pions $n_{\pi}(t,R)$ for all
times $t$ as $n(R)= {\rm const}/R^2$,
where $R$ is the distance from the fireball. 
The `constant' is determinded by the particular freeze-out conditions, namely 
it should be normalized on the real pion density $n_{\rm f}$ at the time of
freeze-out.
This parametrization results from the inequality
$n_{\pi}(t,R)\le {\rm const}/R^2$,
where equality is reached for an expanding system in which the radial particle 
velocities are all equal. 
Quite obviously, owing to the particle velocity dispersion, this 
parametrization is an overestimation of the 
real post-freeze-out pion density at any time $t$. 
Consequently, adopting this model of the pair environment, we 
overestimate the deviation from the Coulomb potential and thus overestimate 
the deviation from the Gamov factor.
A further reduction results from the time evolution of
the mean radii: $R(t)$, which is the distance of the separate pion pair
c.m.s. from the fireball centre, and $r(t)$, which is a classical distance
between these pions. 
Exploiting that reduction, we parametrize the relative
motion, as well as the pair c.m.s. motion, by one variable $r$,
thus eliminating the time dependence. 
After these reductions only one independent variable $r$ is left.

It should be pointed out that the time derivatives can also be
included into our consideration. From the 
$r$-parametrization, we obtain a proportionality of the time derivative
to the pion--pion relative velocity $v_{\rm rel}$, namely  
$\frac{\partial}{\partial t} =v_{\rm rel}\, \frac{\partial}{\partial r} $.
Since we are interested only in the small relative pion momenta
this derivative can be neglected for a first-order estimate.  

So, we reduce the problem to a stationary one where, in contrast to 
the first scenario, the pion pair moves after freeze-out in a non-uniform
environment. 
Effectively, this means that we consider the problem in the two-pion
rest frame, where the relative radial motion of the particles that 
create the environment is slow enough with respect to the radial expansion.
Practically it allows us to consider the stationary Schr\"{o}dinger 
equation instead of the time-dependent one.
We show that the main contribution to the behaviour of the correction factor
comes from the behaviour of the potential at the large distances that separate
interacting particles. On the other hand, the interacting particles
are separated by large distances when they are far enough from the fireball
(this statement is a basis of the $r$-parametrization).
Hence, because of a very fast decrease of the density with respect to the 
distance from the fireball $R$, namely
$n(R)= {\rm const}/R^2$,
the density of secondaries is small or even negligible at these distances.
At this stage of the evolution, the long-range part of the potential is
just the Coulomb one and it is responsible for the behaviour of the correction
factor at small relative momentum $q\le 50~{\rm MeV/c} $. 
That is why the correction factor for this region of relative pion momentum 
practically coincides with the Gamov one. 
The short-range behaviour of the potential (when $r$ is smaller than a mean 
distance between the particles that form the environment),
which is the Coulomb one for point-like pions, provides  also  only small 
deviations of the correction factor $G_{\rm cor}$ from the standard 
Gamov factor $G_0$ for a large pion relative momentum 
$q\ge $50--70~ MeV/c, as can be seen in Figs.~4 and 7.

For the high  ``{\it LHC}'' freeze-out pion density 
$n_{\rm f }=0.25\, {\rm fm}^{-1/3}$ 
the reduction of the correction factor $G_{\rm cor}$, 
as  seen in Fig.~4, to the standard
Gamov factor $G_0$ increases with  decreasing  
$v_{\rm rel}/v_{\rm cm}$, 
the ratio of the relative velocity of the detected pions  
to their c.m. velocity in the fireball rest frame.
When this parameter is much less than unity, the pion pair
promptly escapes the initial high-density region so that 
slow relative motion of the two pions takes place in an approximately empty
spatial region and the distortion of the mutual Coulomb potential is weak.

This is not the case for ``{\it SPS}'' freeze-out conditions, 
as  seen in Fig.~7. 
For the comparetively low pion density $n_{\rm f }=0.03\, {\rm fm}^{-1/3}$ 
the effect of small relative velocity is not pronounced because from the 
very beginning of the freeze-out the pions are not in such a dense 
environment
and even small c.m. velocities are quite sufficient to bring the pair 
promptly into 
regions where the influence of the environment is negligible.

\bigskip

{\bf Acknowledgements}
D.A. acknowledges helpful discussions with G.~Baym, 
D.~Ferenc, V.~Zasenko and would like to express special thanks
to  U.~Heinz, C.~Slotta and J.~Sollfrank who read the manuscript and made 
a number of remarks to improve the presentation.

\bigskip





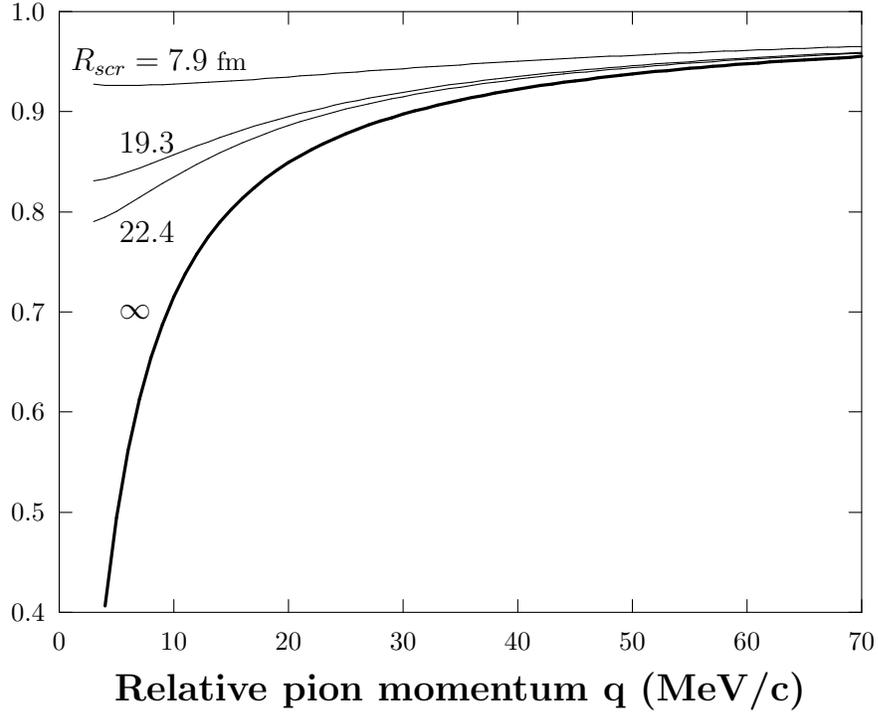
\begin{figure}[htbp]
  \begin{center}
    \leavevmode

\setlength{\unitlength}{0.240900pt}
\ifx\plotpoint\undefined\newsavebox{\plotpoint}\fi
\sbox{\plotpoint}{\rule[-0.200pt]{0.400pt}{0.400pt}}%
\special{em:linewidth 0.4pt}%
\begin{picture}(1500,1080)(0,0)
\font\gnuplot=cmr10 at 10pt
\gnuplot
\put(176,113){\special{em:moveto}}
\put(176,1057){\special{em:lineto}}
\put(176,113){\special{em:moveto}}
\put(196,113){\special{em:lineto}}
\put(1436,113){\special{em:moveto}}
\put(1416,113){\special{em:lineto}}
\put(154,113){\makebox(0,0)[r]{0.4}}
\put(176,270){\special{em:moveto}}
\put(196,270){\special{em:lineto}}
\put(1436,270){\special{em:moveto}}
\put(1416,270){\special{em:lineto}}
\put(154,270){\makebox(0,0)[r]{0.5}}
\put(176,428){\special{em:moveto}}
\put(196,428){\special{em:lineto}}
\put(1436,428){\special{em:moveto}}
\put(1416,428){\special{em:lineto}}
\put(154,428){\makebox(0,0)[r]{0.6}}
\put(176,585){\special{em:moveto}}
\put(196,585){\special{em:lineto}}
\put(1436,585){\special{em:moveto}}
\put(1416,585){\special{em:lineto}}
\put(154,585){\makebox(0,0)[r]{0.7}}
\put(176,742){\special{em:moveto}}
\put(196,742){\special{em:lineto}}
\put(1436,742){\special{em:moveto}}
\put(1416,742){\special{em:lineto}}
\put(154,742){\makebox(0,0)[r]{0.8}}
\put(176,900){\special{em:moveto}}
\put(196,900){\special{em:lineto}}
\put(1436,900){\special{em:moveto}}
\put(1416,900){\special{em:lineto}}
\put(154,900){\makebox(0,0)[r]{0.9}}
\put(176,1057){\special{em:moveto}}
\put(196,1057){\special{em:lineto}}
\put(1436,1057){\special{em:moveto}}
\put(1416,1057){\special{em:lineto}}
\put(154,1057){\makebox(0,0)[r]{1.0}}
\put(176,113){\special{em:moveto}}
\put(176,133){\special{em:lineto}}
\put(176,1057){\special{em:moveto}}
\put(176,1037){\special{em:lineto}}
\put(176,68){\makebox(0,0){0}}
\put(356,113){\special{em:moveto}}
\put(356,133){\special{em:lineto}}
\put(356,1057){\special{em:moveto}}
\put(356,1037){\special{em:lineto}}
\put(356,68){\makebox(0,0){10}}
\put(536,113){\special{em:moveto}}
\put(536,133){\special{em:lineto}}
\put(536,1057){\special{em:moveto}}
\put(536,1037){\special{em:lineto}}
\put(536,68){\makebox(0,0){20}}
\put(716,113){\special{em:moveto}}
\put(716,133){\special{em:lineto}}
\put(716,1057){\special{em:moveto}}
\put(716,1037){\special{em:lineto}}
\put(716,68){\makebox(0,0){30}}
\put(896,113){\special{em:moveto}}
\put(896,133){\special{em:lineto}}
\put(896,1057){\special{em:moveto}}
\put(896,1037){\special{em:lineto}}
\put(896,68){\makebox(0,0){40}}
\put(1076,113){\special{em:moveto}}
\put(1076,133){\special{em:lineto}}
\put(1076,1057){\special{em:moveto}}
\put(1076,1037){\special{em:lineto}}
\put(1076,68){\makebox(0,0){50}}
\put(1256,113){\special{em:moveto}}
\put(1256,133){\special{em:lineto}}
\put(1256,1057){\special{em:moveto}}
\put(1256,1037){\special{em:lineto}}
\put(1256,68){\makebox(0,0){60}}
\put(1436,113){\special{em:moveto}}
\put(1436,133){\special{em:lineto}}
\put(1436,1057){\special{em:moveto}}
\put(1436,1037){\special{em:lineto}}
\put(1436,68){\makebox(0,0){70}}
\put(176,113){\special{em:moveto}}
\put(1436,113){\special{em:lineto}}
\put(1436,1057){\special{em:lineto}}
\put(176,1057){\special{em:lineto}}
\put(176,113){\special{em:lineto}}
\put(806,-10){\makebox(0,0){{\large 
              {\bf Relative pion momentum q (MeV/c)}}}}
\put(194,978){\makebox(0,0)[l]{$R_{scr}=7.9$~fm}}
\put(270,852){\makebox(0,0)[l]{$19.3$}}
\put(270,711){\makebox(0,0)[l]{$22.4$}}
\put(270,585){\makebox(0,0)[l]{$\infty$}}
\put(230,943){\special{em:moveto}}
\put(248,941){\special{em:lineto}}
\put(266,941){\special{em:lineto}}
\put(284,941){\special{em:lineto}}
\put(302,941){\special{em:lineto}}
\put(320,942){\special{em:lineto}}
\put(338,942){\special{em:lineto}}
\put(356,943){\special{em:lineto}}
\put(374,944){\special{em:lineto}}
\put(392,945){\special{em:lineto}}
\put(410,946){\special{em:lineto}}
\put(428,947){\special{em:lineto}}
\put(446,948){\special{em:lineto}}
\put(464,949){\special{em:lineto}}
\put(482,950){\special{em:lineto}}
\put(500,952){\special{em:lineto}}
\put(518,953){\special{em:lineto}}
\put(536,954){\special{em:lineto}}
\put(554,956){\special{em:lineto}}
\put(572,957){\special{em:lineto}}
\put(590,958){\special{em:lineto}}
\put(608,960){\special{em:lineto}}
\put(626,961){\special{em:lineto}}
\put(644,962){\special{em:lineto}}
\put(662,964){\special{em:lineto}}
\put(680,965){\special{em:lineto}}
\put(698,966){\special{em:lineto}}
\put(716,967){\special{em:lineto}}
\put(734,969){\special{em:lineto}}
\put(752,970){\special{em:lineto}}
\put(770,971){\special{em:lineto}}
\put(788,972){\special{em:lineto}}
\put(806,973){\special{em:lineto}}
\put(824,975){\special{em:lineto}}
\put(842,976){\special{em:lineto}}
\put(860,977){\special{em:lineto}}
\put(878,978){\special{em:lineto}}
\put(896,979){\special{em:lineto}}
\put(914,980){\special{em:lineto}}
\put(932,981){\special{em:lineto}}
\put(950,982){\special{em:lineto}}
\put(968,983){\special{em:lineto}}
\put(986,984){\special{em:lineto}}
\put(1004,985){\special{em:lineto}}
\put(1022,986){\special{em:lineto}}
\put(1040,987){\special{em:lineto}}
\put(1058,987){\special{em:lineto}}
\put(1076,988){\special{em:lineto}}
\put(1094,989){\special{em:lineto}}
\put(1112,990){\special{em:lineto}}
\put(1130,991){\special{em:lineto}}
\put(1148,992){\special{em:lineto}}
\put(1166,992){\special{em:lineto}}
\put(1184,993){\special{em:lineto}}
\put(1202,994){\special{em:lineto}}
\put(1220,995){\special{em:lineto}}
\put(1238,995){\special{em:lineto}}
\put(1256,996){\special{em:lineto}}
\put(1274,997){\special{em:lineto}}
\put(1292,997){\special{em:lineto}}
\put(1310,998){\special{em:lineto}}
\put(1328,999){\special{em:lineto}}
\put(1346,999){\special{em:lineto}}
\put(1364,1000){\special{em:lineto}}
\put(1382,1001){\special{em:lineto}}
\put(1400,1001){\special{em:lineto}}
\put(1418,1002){\special{em:lineto}}
\put(1436,1002){\special{em:lineto}}
\put(230,791){\special{em:moveto}}
\put(248,794){\special{em:lineto}}
\put(266,799){\special{em:lineto}}
\put(284,805){\special{em:lineto}}
\put(302,811){\special{em:lineto}}
\put(320,818){\special{em:lineto}}
\put(338,825){\special{em:lineto}}
\put(356,832){\special{em:lineto}}
\put(374,839){\special{em:lineto}}
\put(392,846){\special{em:lineto}}
\put(410,852){\special{em:lineto}}
\put(428,859){\special{em:lineto}}
\put(446,865){\special{em:lineto}}
\put(464,871){\special{em:lineto}}
\put(482,877){\special{em:lineto}}
\put(500,882){\special{em:lineto}}
\put(518,887){\special{em:lineto}}
\put(536,892){\special{em:lineto}}
\put(554,897){\special{em:lineto}}
\put(572,901){\special{em:lineto}}
\put(590,905){\special{em:lineto}}
\put(608,910){\special{em:lineto}}
\put(626,914){\special{em:lineto}}
\put(644,917){\special{em:lineto}}
\put(662,921){\special{em:lineto}}
\put(680,924){\special{em:lineto}}
\put(698,927){\special{em:lineto}}
\put(716,930){\special{em:lineto}}
\put(734,933){\special{em:lineto}}
\put(752,936){\special{em:lineto}}
\put(770,939){\special{em:lineto}}
\put(788,942){\special{em:lineto}}
\put(806,944){\special{em:lineto}}
\put(824,946){\special{em:lineto}}
\put(842,949){\special{em:lineto}}
\put(860,951){\special{em:lineto}}
\put(878,953){\special{em:lineto}}
\put(896,955){\special{em:lineto}}
\put(914,957){\special{em:lineto}}
\put(932,959){\special{em:lineto}}
\put(950,961){\special{em:lineto}}
\put(968,962){\special{em:lineto}}
\put(986,964){\special{em:lineto}}
\put(1004,966){\special{em:lineto}}
\put(1022,967){\special{em:lineto}}
\put(1040,969){\special{em:lineto}}
\put(1058,970){\special{em:lineto}}
\put(1076,972){\special{em:lineto}}
\put(1094,973){\special{em:lineto}}
\put(1112,975){\special{em:lineto}}
\put(1130,976){\special{em:lineto}}
\put(1148,977){\special{em:lineto}}
\put(1166,978){\special{em:lineto}}
\put(1184,980){\special{em:lineto}}
\put(1202,981){\special{em:lineto}}
\put(1220,982){\special{em:lineto}}
\put(1238,983){\special{em:lineto}}
\put(1256,984){\special{em:lineto}}
\put(1274,985){\special{em:lineto}}
\put(1292,986){\special{em:lineto}}
\put(1310,987){\special{em:lineto}}
\put(1328,988){\special{em:lineto}}
\put(1346,989){\special{em:lineto}}
\put(1364,990){\special{em:lineto}}
\put(1382,991){\special{em:lineto}}
\put(1400,991){\special{em:lineto}}
\put(1418,992){\special{em:lineto}}
\put(1436,993){\special{em:lineto}}
\put(230,727){\special{em:moveto}}
\put(248,734){\special{em:lineto}}
\put(266,743){\special{em:lineto}}
\put(284,754){\special{em:lineto}}
\put(302,765){\special{em:lineto}}
\put(320,776){\special{em:lineto}}
\put(338,787){\special{em:lineto}}
\put(356,797){\special{em:lineto}}
\put(374,807){\special{em:lineto}}
\put(392,817){\special{em:lineto}}
\put(410,826){\special{em:lineto}}
\put(428,835){\special{em:lineto}}
\put(446,843){\special{em:lineto}}
\put(464,851){\special{em:lineto}}
\put(482,858){\special{em:lineto}}
\put(500,865){\special{em:lineto}}
\put(518,872){\special{em:lineto}}
\put(536,878){\special{em:lineto}}
\put(554,884){\special{em:lineto}}
\put(572,889){\special{em:lineto}}
\put(590,894){\special{em:lineto}}
\put(608,899){\special{em:lineto}}
\put(626,904){\special{em:lineto}}
\put(644,908){\special{em:lineto}}
\put(662,912){\special{em:lineto}}
\put(680,916){\special{em:lineto}}
\put(698,920){\special{em:lineto}}
\put(716,923){\special{em:lineto}}
\put(734,927){\special{em:lineto}}
\put(752,930){\special{em:lineto}}
\put(770,933){\special{em:lineto}}
\put(788,936){\special{em:lineto}}
\put(806,938){\special{em:lineto}}
\put(824,941){\special{em:lineto}}
\put(842,944){\special{em:lineto}}
\put(860,946){\special{em:lineto}}
\put(878,948){\special{em:lineto}}
\put(896,951){\special{em:lineto}}
\put(914,953){\special{em:lineto}}
\put(932,955){\special{em:lineto}}
\put(950,957){\special{em:lineto}}
\put(968,959){\special{em:lineto}}
\put(986,961){\special{em:lineto}}
\put(1004,962){\special{em:lineto}}
\put(1022,964){\special{em:lineto}}
\put(1040,966){\special{em:lineto}}
\put(1058,967){\special{em:lineto}}
\put(1076,969){\special{em:lineto}}
\put(1094,970){\special{em:lineto}}
\put(1112,972){\special{em:lineto}}
\put(1130,973){\special{em:lineto}}
\put(1148,975){\special{em:lineto}}
\put(1166,976){\special{em:lineto}}
\put(1184,977){\special{em:lineto}}
\put(1202,978){\special{em:lineto}}
\put(1220,980){\special{em:lineto}}
\put(1238,981){\special{em:lineto}}
\put(1256,982){\special{em:lineto}}
\put(1274,983){\special{em:lineto}}
\put(1292,984){\special{em:lineto}}
\put(1310,985){\special{em:lineto}}
\put(1328,986){\special{em:lineto}}
\put(1346,987){\special{em:lineto}}
\put(1364,988){\special{em:lineto}}
\put(1382,989){\special{em:lineto}}
\put(1400,990){\special{em:lineto}}
\put(1418,991){\special{em:lineto}}
\put(1436,991){\special{em:lineto}}
\sbox{\plotpoint}{\rule[-0.600pt]{1.200pt}{1.200pt}}%
\special{em:linewidth 1.2pt}%
\put(248,123){\special{em:moveto}}
\put(266,262){\special{em:lineto}}
\put(284,367){\special{em:lineto}}
\put(302,448){\special{em:lineto}}
\put(320,513){\special{em:lineto}}
\put(338,565){\special{em:lineto}}
\put(356,609){\special{em:lineto}}
\put(374,645){\special{em:lineto}}
\put(392,676){\special{em:lineto}}
\put(410,703){\special{em:lineto}}
\put(428,726){\special{em:lineto}}
\put(446,746){\special{em:lineto}}
\put(464,764){\special{em:lineto}}
\put(482,780){\special{em:lineto}}
\put(500,795){\special{em:lineto}}
\put(518,808){\special{em:lineto}}
\put(536,820){\special{em:lineto}}
\put(554,830){\special{em:lineto}}
\put(572,840){\special{em:lineto}}
\put(590,849){\special{em:lineto}}
\put(608,857){\special{em:lineto}}
\put(626,865){\special{em:lineto}}
\put(644,872){\special{em:lineto}}
\put(662,879){\special{em:lineto}}
\put(680,885){\special{em:lineto}}
\put(698,890){\special{em:lineto}}
\put(716,896){\special{em:lineto}}
\put(734,901){\special{em:lineto}}
\put(752,905){\special{em:lineto}}
\put(770,910){\special{em:lineto}}
\put(788,914){\special{em:lineto}}
\put(806,918){\special{em:lineto}}
\put(824,922){\special{em:lineto}}
\put(842,925){\special{em:lineto}}
\put(860,929){\special{em:lineto}}
\put(878,932){\special{em:lineto}}
\put(896,935){\special{em:lineto}}
\put(914,938){\special{em:lineto}}
\put(932,941){\special{em:lineto}}
\put(950,943){\special{em:lineto}}
\put(968,946){\special{em:lineto}}
\put(986,948){\special{em:lineto}}
\put(1004,950){\special{em:lineto}}
\put(1022,953){\special{em:lineto}}
\put(1040,955){\special{em:lineto}}
\put(1058,957){\special{em:lineto}}
\put(1076,959){\special{em:lineto}}
\put(1094,961){\special{em:lineto}}
\put(1112,963){\special{em:lineto}}
\put(1130,964){\special{em:lineto}}
\put(1148,966){\special{em:lineto}}
\put(1166,968){\special{em:lineto}}
\put(1184,969){\special{em:lineto}}
\put(1202,971){\special{em:lineto}}
\put(1220,972){\special{em:lineto}}
\put(1238,974){\special{em:lineto}}
\put(1256,975){\special{em:lineto}}
\put(1274,976){\special{em:lineto}}
\put(1292,978){\special{em:lineto}}
\put(1310,979){\special{em:lineto}}
\put(1328,980){\special{em:lineto}}
\put(1346,981){\special{em:lineto}}
\put(1364,982){\special{em:lineto}}
\put(1382,983){\special{em:lineto}}
\put(1400,984){\special{em:lineto}}
\put(1418,985){\special{em:lineto}}
\put(1436,987){\special{em:lineto}}
\end{picture}

\bigskip
    
    \caption[]{
Correction factor $G_{\rm cor}(q)=\mid \psi_{\bf q} ({\bf r}=0)\mid ^2$
as a function of the relative pion momentum $q=|{\bf q}|$ (MeV/c).
Wave function $\psi_{\bf q} ({\bf r})$ is the solution of the 
Schr\"odinger equation for the screened Coulomb potential (10).
Three curves correspond to different freeze-out conditions:
1) {\it LHC} freeze-out conditions:
$n_{\rm f }=0.25\, {\rm fm}^{-1/3}$, $T_{\rm f} = 190~{\rm MeV}$, 
and $R_{\rm scr}=7.9$ fm (see Eq.~(9));
2) Intermediate case: $R_{\rm scr}=19.3$ fm;
3) {\it SPS} freeze-out conditions: 
$n_{\rm f }=0.03\, {\rm fm}^{-1/3}$, $T_{\rm f} = 190~{\rm MeV}$,
and $R_{\rm scr}=22.4$ fm. \\
Bottom curve is the standard Gamov factor, $R_{\rm scr}=\infty $.}

\label{fig:1}

  \end{center}
\end{figure}


\begin{figure}[htbp]
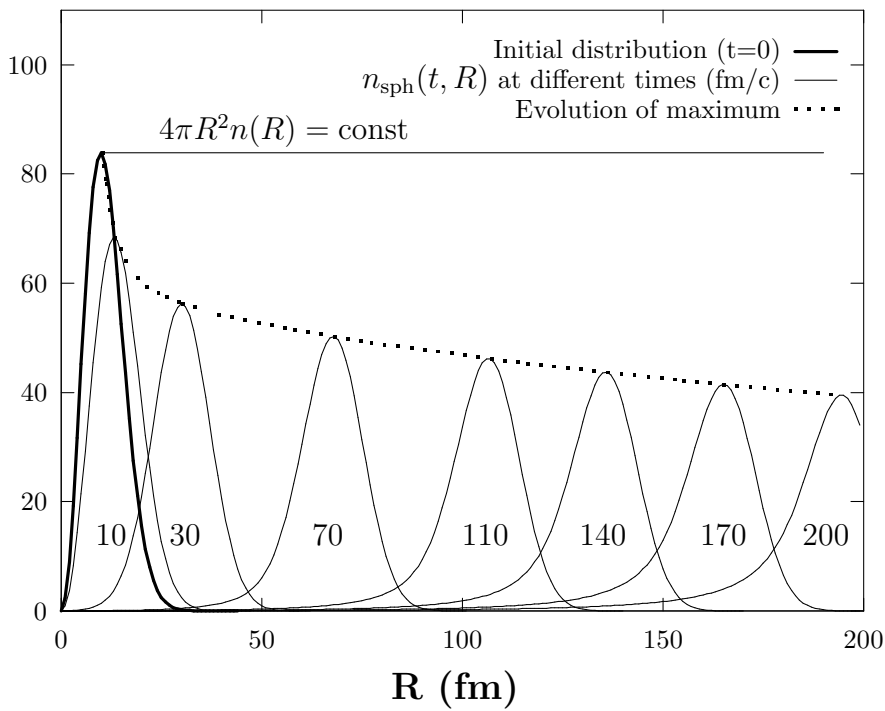

  \begin{center}
    \leavevmode

\setlength{\unitlength}{0.240900pt}
\ifx\plotpoint\undefined\newsavebox{\plotpoint}\fi
\sbox{\plotpoint}{\rule[-0.200pt]{0.400pt}{0.400pt}}%
\special{em:linewidth 0.4pt}%


\bigskip

\caption[]{
Dependence of the spherical pion density
$n_{\rm sph} (t,R) \equiv 4\pi R^2 n(t,R)$ (fm$^{-1}$)
on $R$, where $n(t,R)$ is the number of pions in the unit volume
at time $t$ and at distance $R$ from the center of the fireball
($\int _0^{\infty }dR\, n_{\rm sph} (t,R) =N_{\pi }$).
The density was evaluated for initial data 
associated with {\it SPS} freeze-out conditions:
$n_{\rm f }=0.03\, {\rm fm}^{-1/3}$, $R_{\rm f} = 7.1~{\rm fm}$, 
$T_{\rm f} = 190~{\rm MeV}$.
The spherical pion density (solid Gaussian-like curves) is depicted 
for the times:
1)  $t=0$, initial distribution,
2) $t=10~{\rm fm/c}$,
3) $t=30~{\rm fm/c}$,
4) $t=70~{\rm fm/c}$,
5) $t=110~{\rm fm/c}$,
6) $t=140~{\rm fm/c}$,
7) $t=170~{\rm fm/c}$,
8) $t=200~{\rm fm/c}$.
The dotted curve is the evolution of the maximum of spatial pion distribution.
The horizontal solid line at the top is a
constant spherical density $n_{\rm sph}(t,R)=4\pi R^2 n(R)=$ const 
(fm$^{-1}$).
\label{fig:2} }

  \end{center}
\end{figure}


\begin{figure}[htbp]
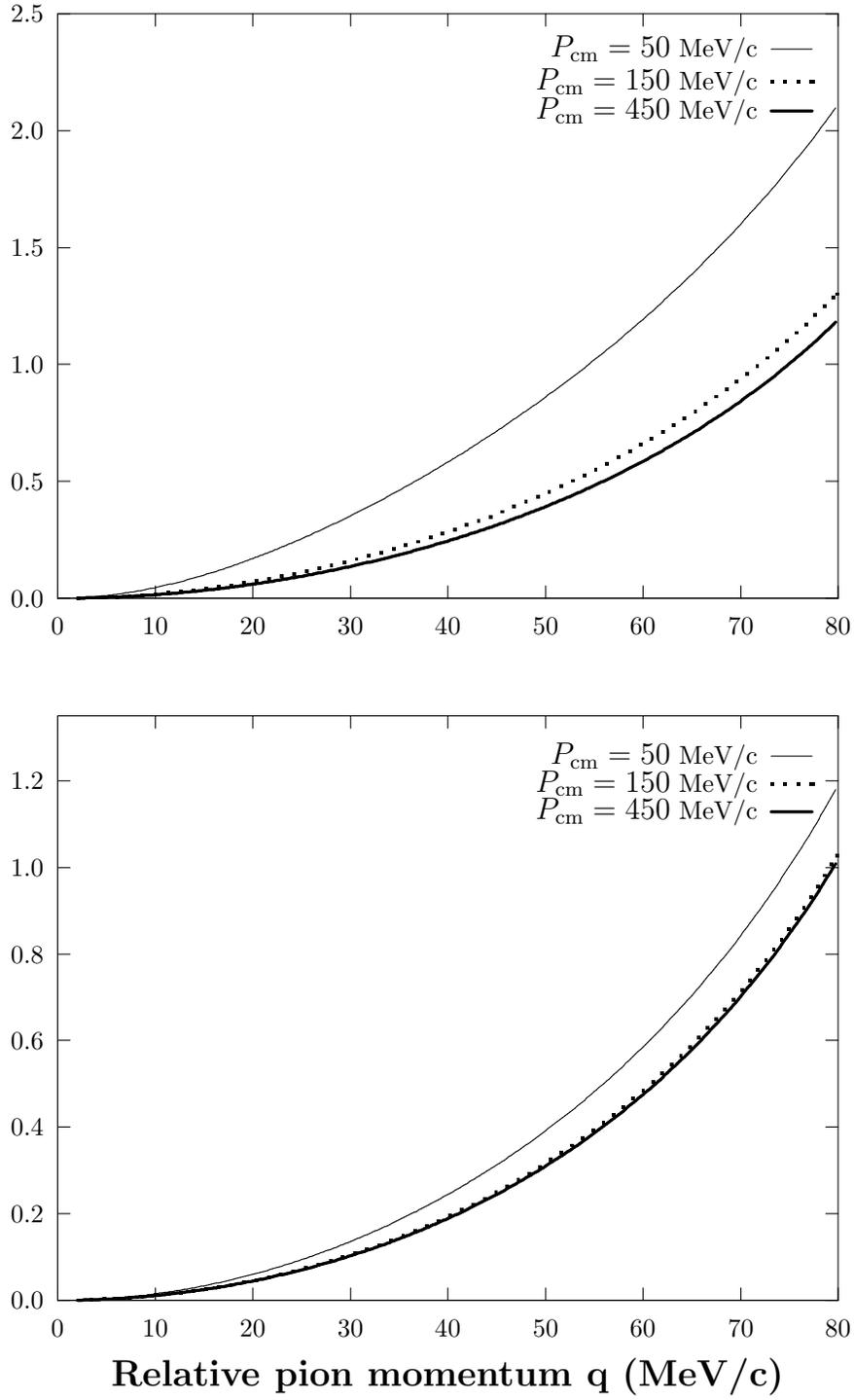

  \begin{center}
    \leavevmode

\setlength{\unitlength}{0.240900pt}
\ifx\plotpoint\undefined\newsavebox{\plotpoint}\fi
\sbox{\plotpoint}{\rule[-0.200pt]{0.400pt}{0.400pt}}%
\special{em:linewidth 0.4pt}%


\bigskip

    \caption[]{
Exponent $b(q)$ (see Eqs.~(31) and (32))\\
a) for the {\it LHC} freeze-out conditions:
$n_{\rm f }=0.25\, {\rm fm}^{-1/3}$, $T_{\rm f} = 190~{\rm MeV}$;\\
b) for {\it SPS} freeze-out conditions:
$n_{\rm f }=0.03\, {\rm fm}^{-1/3}$, $T_{\rm f} = 190~{\rm MeV}$. }

\label{fig:3}

  \end{center}
\end{figure}


\begin{figure}[htbp]
  \begin{center}
    \leavevmode

\setlength{\unitlength}{0.240900pt}
\ifx\plotpoint\undefined\newsavebox{\plotpoint}\fi
\sbox{\plotpoint}{\rule[-0.200pt]{0.400pt}{0.400pt}}%
\special{em:linewidth 0.4pt}%


\bigskip
    
    \caption[]{
The ratio of the standard Gamov factor $G_0$ to the correction factor  
$G_{\rm cor}(q)=$ \\
$\mid \psi_{\bf q} ({\bf r}=0)\mid ^2$  obtained from 
the solution of the Schr\"odinger equation
for the {\it LHC} freeze-out conditions:
$n_{\rm f }=0.25\, {\rm fm}^{-1/3}$, $R_{\rm f} = 7.1~{\rm fm}$, 
$T_{\rm f} = 190~{\rm MeV}$.\\
The dotted curves show the evaluation made using the potential (36),
the solid curves using the improved potential (38).
The curves are drawn for different mean momenta of the pion pair  
$P_{\rm cm}=|{\bf p}_a+{\bf p}_b|/2$ in the fireball frame.
    \label{fig:4} }
  \end{center}
\end{figure}


\begin{figure}[htbp]
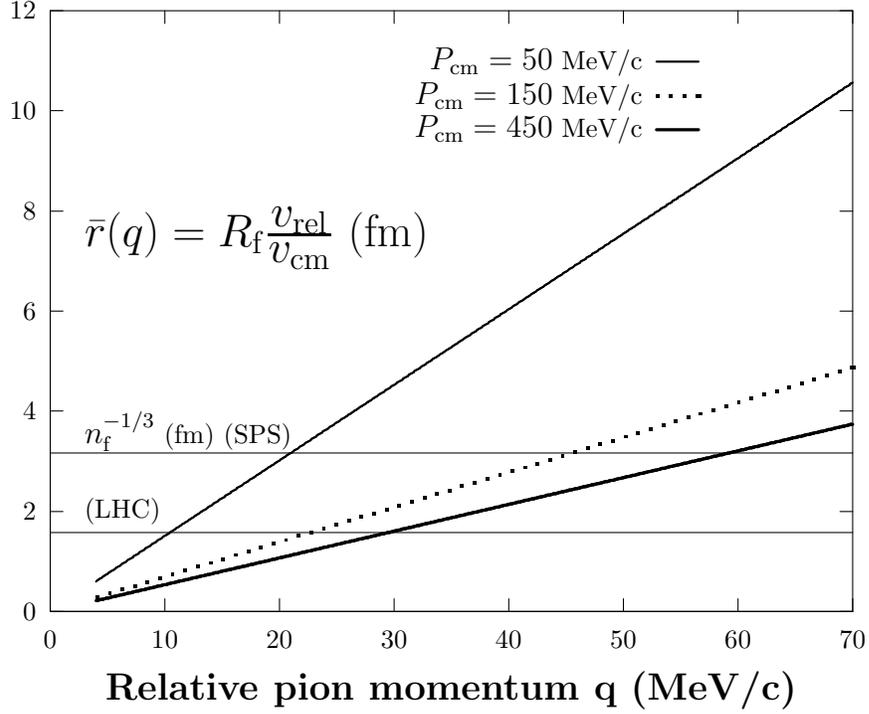

  \begin{center}
    \leavevmode

\setlength{\unitlength}{0.240900pt}
\ifx\plotpoint\undefined\newsavebox{\plotpoint}\fi
\sbox{\plotpoint}{\rule[-0.200pt]{0.400pt}{0.400pt}}%
\special{em:linewidth 0.4pt}%


\bigskip    

    \caption[]{
Dependence of the dynamical freeze-out radius
$\overline{r}(q)= R_{\rm f} \frac{v_{\rm rel}}{v_{\rm cm}}$ (fm)
on the relative pion momentum $q$ (MeV/c). 
Slope lines are drawn for different mean momenta of the pair:
$P_{\rm cm}=50,~150,~450$ MeV/c. 
The mean distance between 
particles at freezeout $a= n_{\rm f}^{-1/3}$ (fm)
are drawn as horizontal lines for {\it SPS} ($a=3.16$\, fm)
and {\it LHC} ($a=1.58$\, fm) freeze-out conditions.}

    \label{fig:5}
  \end{center}
\end{figure}


\begin{figure}[htbp]
  \begin{center}
    \leavevmode

\setlength{\unitlength}{0.240900pt}
\ifx\plotpoint\undefined\newsavebox{\plotpoint}\fi
\sbox{\plotpoint}{\rule[-0.200pt]{0.400pt}{0.400pt}}%
\special{em:linewidth 0.4pt}%
\begin{picture}(1500,1080)(0,0)
\font\gnuplot=cmr10 at 10pt
\gnuplot
\put(176,113){\special{em:moveto}}
\put(1436,113){\special{em:lineto}}
\put(176,113){\special{em:moveto}}
\put(176,1057){\special{em:lineto}}
\put(176,113){\special{em:moveto}}
\put(196,113){\special{em:lineto}}
\put(1436,113){\special{em:moveto}}
\put(1416,113){\special{em:lineto}}
\put(154,113){\makebox(0,0)[r]{0.0}}
\put(176,207){\special{em:moveto}}
\put(196,207){\special{em:lineto}}
\put(1436,207){\special{em:moveto}}
\put(1416,207){\special{em:lineto}}
\put(154,207){\makebox(0,0)[r]{0.1}}
\put(176,302){\special{em:moveto}}
\put(196,302){\special{em:lineto}}
\put(1436,302){\special{em:moveto}}
\put(1416,302){\special{em:lineto}}
\put(154,302){\makebox(0,0)[r]{0.2}}
\put(176,396){\special{em:moveto}}
\put(196,396){\special{em:lineto}}
\put(1436,396){\special{em:moveto}}
\put(1416,396){\special{em:lineto}}
\put(154,396){\makebox(0,0)[r]{0.3}}
\put(176,491){\special{em:moveto}}
\put(196,491){\special{em:lineto}}
\put(1436,491){\special{em:moveto}}
\put(1416,491){\special{em:lineto}}
\put(154,491){\makebox(0,0)[r]{0.4}}
\put(176,585){\special{em:moveto}}
\put(196,585){\special{em:lineto}}
\put(1436,585){\special{em:moveto}}
\put(1416,585){\special{em:lineto}}
\put(154,585){\makebox(0,0)[r]{0.5}}
\put(176,679){\special{em:moveto}}
\put(196,679){\special{em:lineto}}
\put(1436,679){\special{em:moveto}}
\put(1416,679){\special{em:lineto}}
\put(154,679){\makebox(0,0)[r]{0.6}}
\put(176,774){\special{em:moveto}}
\put(196,774){\special{em:lineto}}
\put(1436,774){\special{em:moveto}}
\put(1416,774){\special{em:lineto}}
\put(154,774){\makebox(0,0)[r]{0.7}}
\put(176,868){\special{em:moveto}}
\put(196,868){\special{em:lineto}}
\put(1436,868){\special{em:moveto}}
\put(1416,868){\special{em:lineto}}
\put(154,868){\makebox(0,0)[r]{0.8}}
\put(176,963){\special{em:moveto}}
\put(196,963){\special{em:lineto}}
\put(1436,963){\special{em:moveto}}
\put(1416,963){\special{em:lineto}}
\put(154,963){\makebox(0,0)[r]{0.9}}
\put(176,1057){\special{em:moveto}}
\put(196,1057){\special{em:lineto}}
\put(1436,1057){\special{em:moveto}}
\put(1416,1057){\special{em:lineto}}
\put(154,1057){\makebox(0,0)[r]{1.0}}
\put(176,113){\special{em:moveto}}
\put(176,133){\special{em:lineto}}
\put(176,1057){\special{em:moveto}}
\put(176,1037){\special{em:lineto}}
\put(176,68){\makebox(0,0){0}}
\put(316,113){\special{em:moveto}}
\put(316,133){\special{em:lineto}}
\put(316,1057){\special{em:moveto}}
\put(316,1037){\special{em:lineto}}
\put(316,68){\makebox(0,0){0.5}}
\put(456,113){\special{em:moveto}}
\put(456,133){\special{em:lineto}}
\put(456,1057){\special{em:moveto}}
\put(456,1037){\special{em:lineto}}
\put(456,68){\makebox(0,0){1}}
\put(596,113){\special{em:moveto}}
\put(596,133){\special{em:lineto}}
\put(596,1057){\special{em:moveto}}
\put(596,1037){\special{em:lineto}}
\put(596,68){\makebox(0,0){1.5}}
\put(736,113){\special{em:moveto}}
\put(736,133){\special{em:lineto}}
\put(736,1057){\special{em:moveto}}
\put(736,1037){\special{em:lineto}}
\put(736,68){\makebox(0,0){2}}
\put(876,113){\special{em:moveto}}
\put(876,133){\special{em:lineto}}
\put(876,1057){\special{em:moveto}}
\put(876,1037){\special{em:lineto}}
\put(876,68){\makebox(0,0){2.5}}
\put(1016,113){\special{em:moveto}}
\put(1016,133){\special{em:lineto}}
\put(1016,1057){\special{em:moveto}}
\put(1016,1037){\special{em:lineto}}
\put(1016,68){\makebox(0,0){3}}
\put(1156,113){\special{em:moveto}}
\put(1156,133){\special{em:lineto}}
\put(1156,1057){\special{em:moveto}}
\put(1156,1037){\special{em:lineto}}
\put(1156,68){\makebox(0,0){3.5}}
\put(1296,113){\special{em:moveto}}
\put(1296,133){\special{em:lineto}}
\put(1296,1057){\special{em:moveto}}
\put(1296,1037){\special{em:lineto}}
\put(1296,68){\makebox(0,0){4}}
\put(1436,113){\special{em:moveto}}
\put(1436,133){\special{em:lineto}}
\put(1436,1057){\special{em:moveto}}
\put(1436,1037){\special{em:lineto}}
\put(1436,68){\makebox(0,0){4.5}}
\put(176,113){\special{em:moveto}}
\put(1436,113){\special{em:lineto}}
\put(1436,1057){\special{em:lineto}}
\put(176,1057){\special{em:lineto}}
\put(176,113){\special{em:lineto}}
\put(806,-10){\makebox(0,0){\Large {\bf x}}}
\put(1306,950){\makebox(0,0)[r]{\large s(x)}}
\put(1328,950){\special{em:moveto}}
\put(1394,950){\special{em:lineto}}
\put(176,1057){\special{em:moveto}}
\put(189,1057){\special{em:lineto}}
\put(201,1057){\special{em:lineto}}
\put(214,1056){\special{em:lineto}}
\put(227,1056){\special{em:lineto}}
\put(240,1056){\special{em:lineto}}
\put(252,1056){\special{em:lineto}}
\put(265,1056){\special{em:lineto}}
\put(278,1056){\special{em:lineto}}
\put(291,1055){\special{em:lineto}}
\put(303,1055){\special{em:lineto}}
\put(316,1055){\special{em:lineto}}
\put(329,1054){\special{em:lineto}}
\put(341,1054){\special{em:lineto}}
\put(354,1053){\special{em:lineto}}
\put(367,1052){\special{em:lineto}}
\put(380,1051){\special{em:lineto}}
\put(392,1050){\special{em:lineto}}
\put(405,1049){\special{em:lineto}}
\put(418,1047){\special{em:lineto}}
\put(431,1045){\special{em:lineto}}
\put(443,1043){\special{em:lineto}}
\put(456,1040){\special{em:lineto}}
\put(469,1037){\special{em:lineto}}
\put(481,1033){\special{em:lineto}}
\put(494,1028){\special{em:lineto}}
\put(507,1023){\special{em:lineto}}
\put(520,1016){\special{em:lineto}}
\put(532,1008){\special{em:lineto}}
\put(545,999){\special{em:lineto}}
\put(558,988){\special{em:lineto}}
\put(571,976){\special{em:lineto}}
\put(583,961){\special{em:lineto}}
\put(596,944){\special{em:lineto}}
\put(609,925){\special{em:lineto}}
\put(621,903){\special{em:lineto}}
\put(634,878){\special{em:lineto}}
\put(647,850){\special{em:lineto}}
\put(660,820){\special{em:lineto}}
\put(672,786){\special{em:lineto}}
\put(685,749){\special{em:lineto}}
\put(698,711){\special{em:lineto}}
\put(711,670){\special{em:lineto}}
\put(723,628){\special{em:lineto}}
\put(736,585){\special{em:lineto}}
\put(749,542){\special{em:lineto}}
\put(761,500){\special{em:lineto}}
\put(774,459){\special{em:lineto}}
\put(787,421){\special{em:lineto}}
\put(800,384){\special{em:lineto}}
\put(812,350){\special{em:lineto}}
\put(825,320){\special{em:lineto}}
\put(838,292){\special{em:lineto}}
\put(851,267){\special{em:lineto}}
\put(863,245){\special{em:lineto}}
\put(876,226){\special{em:lineto}}
\put(889,209){\special{em:lineto}}
\put(901,194){\special{em:lineto}}
\put(914,182){\special{em:lineto}}
\put(927,171){\special{em:lineto}}
\put(940,162){\special{em:lineto}}
\put(952,154){\special{em:lineto}}
\put(965,147){\special{em:lineto}}
\put(978,142){\special{em:lineto}}
\put(991,137){\special{em:lineto}}
\put(1003,133){\special{em:lineto}}
\put(1016,130){\special{em:lineto}}
\put(1029,127){\special{em:lineto}}
\put(1041,125){\special{em:lineto}}
\put(1054,123){\special{em:lineto}}
\put(1067,121){\special{em:lineto}}
\put(1080,120){\special{em:lineto}}
\put(1092,119){\special{em:lineto}}
\put(1105,118){\special{em:lineto}}
\put(1118,117){\special{em:lineto}}
\put(1131,116){\special{em:lineto}}
\put(1143,116){\special{em:lineto}}
\put(1156,115){\special{em:lineto}}
\put(1169,115){\special{em:lineto}}
\put(1181,115){\special{em:lineto}}
\put(1194,114){\special{em:lineto}}
\put(1207,114){\special{em:lineto}}
\put(1220,114){\special{em:lineto}}
\put(1232,114){\special{em:lineto}}
\put(1245,114){\special{em:lineto}}
\put(1258,114){\special{em:lineto}}
\put(1271,113){\special{em:lineto}}
\put(1283,113){\special{em:lineto}}
\put(1296,113){\special{em:lineto}}
\put(1309,113){\special{em:lineto}}
\put(1321,113){\special{em:lineto}}
\put(1334,113){\special{em:lineto}}
\put(1347,113){\special{em:lineto}}
\put(1360,113){\special{em:lineto}}
\put(1372,113){\special{em:lineto}}
\put(1385,113){\special{em:lineto}}
\put(1398,113){\special{em:lineto}}
\put(1411,113){\special{em:lineto}}
\put(1423,113){\special{em:lineto}}
\put(1436,113){\special{em:lineto}}
\sbox{\plotpoint}{\rule[-0.500pt]{1.000pt}{1.000pt}}%
\special{em:linewidth 1.0pt}%
\put(1306,890){\makebox(0,0)[r]{\large 1-s(x)}}
\multiput(1328,890)(20.756,0.000){4}{\usebox{\plotpoint}}
\put(1394,890){\usebox{\plotpoint}}
\put(176,113){\usebox{\plotpoint}}
\put(176.00,113.00){\usebox{\plotpoint}}
\put(196.76,113.00){\usebox{\plotpoint}}
\multiput(201,113)(20.694,1.592){0}{\usebox{\plotpoint}}
\put(217.47,114.00){\usebox{\plotpoint}}
\put(238.23,114.00){\usebox{\plotpoint}}
\multiput(240,114)(20.756,0.000){0}{\usebox{\plotpoint}}
\put(258.98,114.00){\usebox{\plotpoint}}
\multiput(265,114)(20.756,0.000){0}{\usebox{\plotpoint}}
\put(279.73,114.13){\usebox{\plotpoint}}
\put(300.46,115.00){\usebox{\plotpoint}}
\multiput(303,115)(20.756,0.000){0}{\usebox{\plotpoint}}
\put(321.20,115.40){\usebox{\plotpoint}}
\multiput(329,116)(20.756,0.000){0}{\usebox{\plotpoint}}
\put(341.93,116.07){\usebox{\plotpoint}}
\put(362.62,117.66){\usebox{\plotpoint}}
\multiput(367,118)(20.694,1.592){0}{\usebox{\plotpoint}}
\put(383.31,119.28){\usebox{\plotpoint}}
\put(404.00,120.92){\usebox{\plotpoint}}
\multiput(405,121)(20.514,3.156){0}{\usebox{\plotpoint}}
\put(424.53,124.00){\usebox{\plotpoint}}
\multiput(431,125)(20.473,3.412){0}{\usebox{\plotpoint}}
\put(444.99,127.46){\usebox{\plotpoint}}
\put(465.21,132.13){\usebox{\plotpoint}}
\multiput(469,133)(19.690,6.563){0}{\usebox{\plotpoint}}
\put(484.94,138.51){\usebox{\plotpoint}}
\put(504.31,145.97){\usebox{\plotpoint}}
\multiput(507,147)(18.275,9.840){0}{\usebox{\plotpoint}}
\put(522.59,155.72){\usebox{\plotpoint}}
\put(539.76,167.37){\usebox{\plotpoint}}
\put(555.98,180.29){\usebox{\plotpoint}}
\multiput(558,182)(15.251,14.078){0}{\usebox{\plotpoint}}
\put(571.26,194.33){\usebox{\plotpoint}}
\put(584.19,210.56){\usebox{\plotpoint}}
\multiput(596,226)(11.720,17.130){2}{\usebox{\plotpoint}}
\put(618.49,262.39){\usebox{\plotpoint}}
\put(628.15,280.76){\usebox{\plotpoint}}
\multiput(634,292)(8.740,18.825){2}{\usebox{\plotpoint}}
\put(654.44,337.18){\usebox{\plotpoint}}
\multiput(660,350)(6.908,19.572){2}{\usebox{\plotpoint}}
\multiput(672,384)(6.880,19.582){2}{\usebox{\plotpoint}}
\multiput(685,421)(6.718,19.638){2}{\usebox{\plotpoint}}
\multiput(698,459)(6.273,19.785){2}{\usebox{\plotpoint}}
\multiput(711,500)(5.702,19.957){2}{\usebox{\plotpoint}}
\multiput(723,542)(6.006,19.867){2}{\usebox{\plotpoint}}
\multiput(736,585)(6.006,19.867){2}{\usebox{\plotpoint}}
\multiput(749,628)(5.702,19.957){2}{\usebox{\plotpoint}}
\multiput(761,670)(6.273,19.785){2}{\usebox{\plotpoint}}
\multiput(774,711)(6.718,19.638){2}{\usebox{\plotpoint}}
\multiput(787,749)(6.880,19.582){2}{\usebox{\plotpoint}}
\multiput(800,786)(6.908,19.572){2}{\usebox{\plotpoint}}
\multiput(812,820)(8.253,19.044){2}{\usebox{\plotpoint}}
\put(833.27,867.81){\usebox{\plotpoint}}
\put(842.39,886.44){\usebox{\plotpoint}}
\multiput(851,903)(9.939,18.221){2}{\usebox{\plotpoint}}
\put(873.47,940.30){\usebox{\plotpoint}}
\put(885.89,956.93){\usebox{\plotpoint}}
\put(898.77,973.21){\usebox{\plotpoint}}
\put(913.62,987.65){\usebox{\plotpoint}}
\multiput(914,988)(15.844,13.407){0}{\usebox{\plotpoint}}
\put(929.64,1000.83){\usebox{\plotpoint}}
\put(946.79,1012.52){\usebox{\plotpoint}}
\put(964.76,1022.87){\usebox{\plotpoint}}
\multiput(965,1023)(19.372,7.451){0}{\usebox{\plotpoint}}
\put(984.12,1030.35){\usebox{\plotpoint}}
\multiput(991,1033)(19.690,6.563){0}{\usebox{\plotpoint}}
\put(1003.71,1037.16){\usebox{\plotpoint}}
\put(1023.94,1041.83){\usebox{\plotpoint}}
\multiput(1029,1043)(20.473,3.412){0}{\usebox{\plotpoint}}
\put(1044.35,1045.52){\usebox{\plotpoint}}
\put(1064.87,1048.67){\usebox{\plotpoint}}
\multiput(1067,1049)(20.694,1.592){0}{\usebox{\plotpoint}}
\put(1085.54,1050.46){\usebox{\plotpoint}}
\multiput(1092,1051)(20.694,1.592){0}{\usebox{\plotpoint}}
\put(1106.23,1052.09){\usebox{\plotpoint}}
\put(1126.93,1053.69){\usebox{\plotpoint}}
\multiput(1131,1054)(20.756,0.000){0}{\usebox{\plotpoint}}
\put(1147.66,1054.36){\usebox{\plotpoint}}
\put(1168.39,1055.00){\usebox{\plotpoint}}
\multiput(1169,1055)(20.756,0.000){0}{\usebox{\plotpoint}}
\put(1189.12,1055.62){\usebox{\plotpoint}}
\multiput(1194,1056)(20.756,0.000){0}{\usebox{\plotpoint}}
\put(1209.86,1056.00){\usebox{\plotpoint}}
\put(1230.61,1056.00){\usebox{\plotpoint}}
\multiput(1232,1056)(20.756,0.000){0}{\usebox{\plotpoint}}
\put(1251.37,1056.00){\usebox{\plotpoint}}
\multiput(1258,1056)(20.694,1.592){0}{\usebox{\plotpoint}}
\put(1272.09,1057.00){\usebox{\plotpoint}}
\put(1292.84,1057.00){\usebox{\plotpoint}}
\multiput(1296,1057)(20.756,0.000){0}{\usebox{\plotpoint}}
\put(1313.60,1057.00){\usebox{\plotpoint}}
\multiput(1321,1057)(20.756,0.000){0}{\usebox{\plotpoint}}
\put(1334.35,1057.00){\usebox{\plotpoint}}
\put(1355.11,1057.00){\usebox{\plotpoint}}
\multiput(1360,1057)(20.756,0.000){0}{\usebox{\plotpoint}}
\put(1375.86,1057.00){\usebox{\plotpoint}}
\put(1396.62,1057.00){\usebox{\plotpoint}}
\multiput(1398,1057)(20.756,0.000){0}{\usebox{\plotpoint}}
\put(1417.38,1057.00){\usebox{\plotpoint}}
\multiput(1423,1057)(20.756,0.000){0}{\usebox{\plotpoint}}
\end{picture}

\bigskip

    \caption{
The solid curve is a smooth potential switcher
$s(x)= \frac{1}{2} - \frac{1}{2} {\rm tanh}\, [2(x-2)]$.
The dotted curve is an alternative potential switcher $1-s(x)$.}

    \label{fig:6}
  \end{center}
\end{figure}
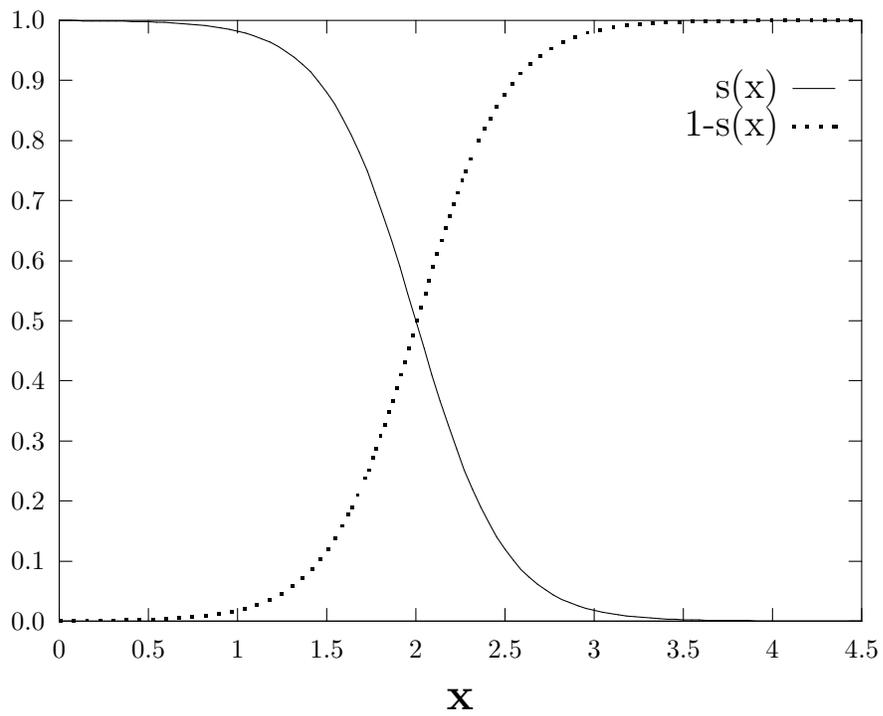


\begin{figure}[htbp]
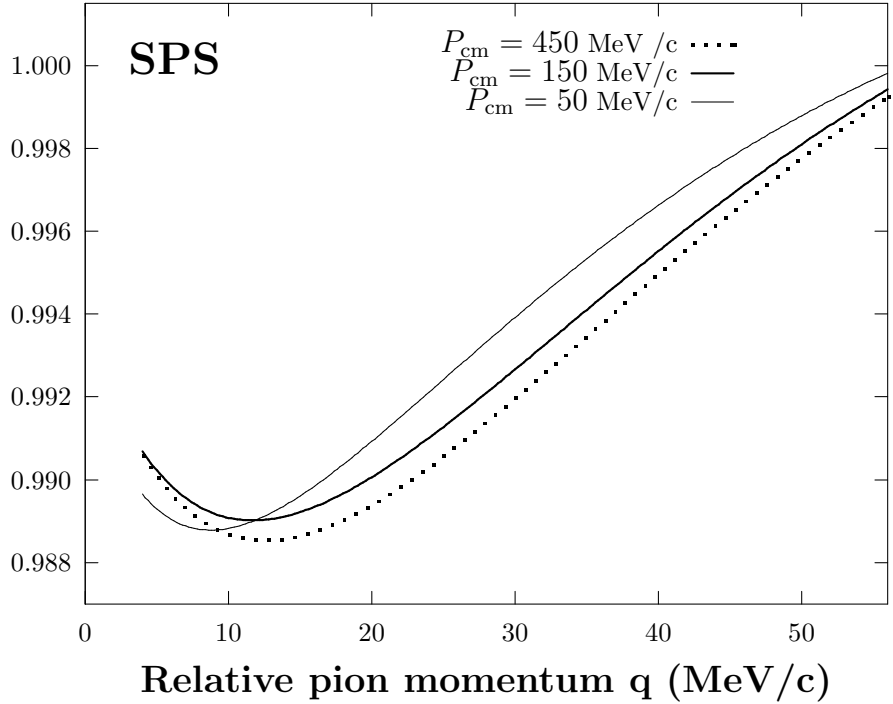

  \begin{center}
    \leavevmode

\setlength{\unitlength}{0.240900pt}
\ifx\plotpoint\undefined\newsavebox{\plotpoint}\fi
\sbox{\plotpoint}{\rule[-0.200pt]{0.400pt}{0.400pt}}%
\special{em:linewidth 0.4pt}%


\bigskip

    \caption[]{
The ratio of the standard Gamov factor $G_0$ to the correction factor
$G_{\rm cor}(q)=$ \\
$\mid \psi_{\bf q} ({\bf r}=0)\mid ^2$ obtained from the 
solution of the Schr\"odinger equation, using the improved 
potential (38) for {\it SPS} freeze-out conditions:
$n_{\rm f }=0.03\, {\rm fm}^{-1/3}$, $R_{\rm f} = 7.1~{\rm fm}$, 
$T_{\rm f} = 190~{\rm MeV}$.
The curves are drawn for different mean momenta of the pion pair  
$P_{\rm cm}=|{\bf p}_a+{\bf p}_b|/2$ in the fireball frame.
    \label{fig:7} }
  \end{center}
\end{figure}

\end{document}